\documentclass[12pt]{article}

\usepackage{graphics}
\usepackage{amssymb} 

\def\afterthmseparator{.}
\makeatletter
\renewcommand{\@begintheorem}[2]{\trivlist
      \item[\hskip \labelsep{\bf #1\ #2\unskip\afterthmseparator}]\em}
\renewcommand{\@opargbegintheorem}[3]{\trivlist
      \item[\hskip \labelsep{\bf #1\ #2\ (#3)\unskip\afterthmseparator}]\em}
\makeatother
\newenvironment{bfenumerate}%
{\begin{enumerate}}%
{\end{enumerate}}
\newcounter{oq}
\newcommand{\que}{\refstepcounter{oq}\par{\bf \theoq.}~}
\newtheorem{theorem}{Theorem}
\newtheorem{lemma}[theorem]{Lemma}

\newcommand{\bull}{\mbox{$\;\;\;$\vrule height .9ex width .8ex depth -.1ex}}

\newenvironment{proof}{\par\smallbreak\noindent{\bf Proof.~}}
{\unskip\nobreak\hfill \bull \par\medbreak}
\newcommand{\setdef}[2]{\left\{ \hspace{0.5mm} #1 : \hspace{0.5mm} #2 \right\}}

\newcommand{\of}[1]{\left( #1 \right)}

\newcommand{\EE}[1]{ {\mathbb E} \left[ #1 \right] }
\newcommand{\obf}[1]{\mathit{obf}(#1)}
\newcommand{\shift}[1]{\mathit{shift}(#1)}
\newcommand{\eps}{\epsilon}

\title{On the Obfuscation Complexity\\ of Planar Graphs}

\author{Oleg Verbitsky%
\thanks{Supported by an Alexander von Humboldt return fellowship.}\\[4mm]
IAPMM, Lviv 79060, Ukraine}

\date{}

\begin{document}

\sloppy

\maketitle

\begin{abstract}
Being motivated by John Tantalo's Planarity Game, we 
consider straight line plane drawings of a planar graph $G$ with edge crossings 
and wonder how obfuscated such drawings can be.
We define $\obf G$, the \emph{obfuscation complexity} of $G$, 
to be the maximum number of edge crossings in a drawing of $G$. 
Relating $\obf G$ to the distribution of vertex degrees in $G$, we show an 
efficient way of constructing a drawing of $G$ with at least $\obf G/3$ edge 
crossings. We prove bounds $(\delta(G)^2/24-o(1))\,n^2\le\obf G<3\,n^2$ for an $n$-vertex
planar graph $G$ with minimum vertex degree $\delta(G)\ge2$.

The \emph{shift complexity} of $G$, denoted by $\shift G$, 
is the minimum number of vertex shifts sufficient to eliminate all edge 
crossings in an arbitrarily obfuscated drawing of $G$ (after shifting a 
vertex, all incident edges are supposed to be redrawn correspondingly).
If $\delta(G)\ge3$, then $\shift G$ is linear in the number of 
vertices due to the known fact that the matching number of $G$ is linear. 
However, in the case $\delta(G)\ge2$ we notice that $\shift G$ can be linear 
even if the matching number is bounded.
As for computational complexity, we show that, given a drawing $D$ of a planar graph,
it is NP-hard to find an optimum sequence of shifts making $D$ crossing-free.
\end{abstract}

\section{Introduction}

This note is inspired by John Tantalo's Planarity Game \cite{Tan}
(another implementation is available at \cite{alcox}).
An instance of the game is a straight line drawing of a planar graph with 
many edge crossings. In a move the player is able to shift one vertex of the 
graph to a new position; the incident edges will be redrawn correspondingly. 
The objective is to achieve a crossing-free drawing in a possibly smaller 
number of moves.

Let us fix some relevant terminology. By a \emph{drawing} we will always 
mean a straight line plane drawing of a graph where no vertex is an inner 
point of any edge. An \emph{edge crossing} in a drawing $D$ is a pair of
edges having a common inner point. The number of edge crossings in $D$ will 
be denoted by $\obf D$. We define the \emph{obfuscation complexity} of
a graph $G$ to be the maximum $\obf D$ over all drawings $D$ of $G$. This graph
parameter will be denoted by $\obf G$.

Given a drawing $D$ of a planar graph $G$, let $\shift D$ denote the minimum 
number of vertex shifts making $D$ crossing-free. The \emph{shift 
complexity} of $G$, denoted by $\shift G$, is the maximum $\shift D$ over all 
drawings of $G$.

Our aim is a combinatorial and a complexity-theoretic 
analysis of the Planarity Game from the
standpoint of a game designer. The latter should definitely have a library of
planar graphs $G$ with large $\shift G$. Generation of planar graphs with large
$\obf G$ is also of interest. Though large obfuscation complexity does not 
imply large shift complexity (see discussion in Section \ref{s:concl}.\ref{oq:obfvsshift}), 
the designer can at least expect that a large $\obf D$ will be 
a psychological obstacle for a player to play optimally on~$D$.

A result of direct relevance to the topic is obtained by
Pach and Tardos \cite{PTa}. Somewhat surprisingly, they prove
that even cycles have large shift complexity, namely,
$
n - O((n\log n)^{2/3})\le\shift{C_n}\le n-\lfloor\sqrt n\rfloor
$.

We first address the obfuscation complexity.
In Section \ref{s:obf} we relate this parameter of a graph to the 
distribution of its vertex degrees. This gives us an
efficient way of constructing a drawing $D$ of a given graph $G$
so that $\obf D\ge\obf G/3$. 
As another consequence, we prove that $\obf G\ge(\delta(G)^2/24-o(1))n^2$
for an $n$-vertex planar graph with minimum vertex degree $\delta(G)\ge2$.
On the other hand, we prove an upper bound $\obf G<3\,n^2$.
In Section \ref{s:shift} we discuss the relationship between the shift complexity of a 
planar graph and its matching number. We also show that the shift complexity
of a drawing is NP-hard to compute.
Section \ref{s:concl} contains concluding remarks and questions.

{\bf Related work.}
Investigation of the parameter $\shift G$ is well motivated from a graph drawing
perspective. Several results were obtained in this area independently of our work
and appeared in \cite{Goaos,SWo,Bose} soon after the present note was submitted
to the journal. The Planarity Game is also mentioned in \cite{Goaos,SWo}
as a source of motivation.

Goaos et al.\ \cite{Goaos} independently prove that computing $\shift D$ for
a given drawing $D$ is an NP-hard problem, the same result as stated in our
Theorem \ref{thm:hard}. They use a different reduction, allowing them to show
that $\shift D$ is even hard to approximate. Our reduction has another
advantage: It shows that it is NP-hard to untangle even drawings of as simple
graphs as matchings.

Spillner and Wolff \cite{SWo} and Bose et al.\ \cite{Bose} obtain
general upper bounds for $\shift G$, which quantitatively improve
the classical Wagner-F\'ary-Stein theorem (cf.\ Theorem \ref{thm:WFS} in Section 
\ref{s:shift}). The stronger of their bounds \cite{Bose} claims that
$\shift G\le n-\sqrt[4]{n/9}$ for any planar $G$. Even better bounds are
established for trees \cite{Goaos} and outerplanar graphs \cite{SWo}.
The series of papers \cite{Goaos,SWo,Bose} gives also lower bounds on
the variant of $\shift G$ for a broader notion of a ``bad drawing''.

{\bf Notation.}
We reserve $n$ and $m$ for, respectively, the number of vertices
and the number of edges in a graph under consideration. We use the standard
notation $K_n$, $K_{s,t}$, and $C_n$ for, respectively, complete graphs, complete
bipartite graphs, and cycles. The vertex set of a graph $G$ will be denoted by
$V(G)$. By $kG$ we mean the disjoint union of $k$ copies of $G$.
The number of edges emanating from a vertex $v$ is called the \emph{degree} 
of $v$ and denoted by $\deg v$. The \emph{minimum degree} of a graph $G$
is defined by $\delta(G)=\min_{v\in V(G)}\deg v$. A set of pairwise non-adjacent
vertices (resp., edges) is called an \emph{independent set}
(resp., a \emph{matching}). The maximum cardinality of an independent set
(resp., a matching) in a graph $G$ is denoted by $\alpha(G)$ (resp., $\nu(G)$)
and called the \emph{independence number} (resp., the \emph{matching
number}) of~$G$. A graph is \emph{$k$-connected} if it stays connected
after removal of any $k-1$ vertices.

\section{Estimation of the obfuscation complexity}\label{s:obf}

Note that $\obf G$ is well defined for an arbitrary, not necessary planar graph
$G$. As a warm-up, consider a few examples.

\begin{description}
\item{$\obf{K_n}={n\choose 4}$.}
Indeed, let $D$ be a drawing of $K_n$. $\obf D$ is computable as follows.
We start with the initial value $0$ and, tracing through all pairs $\{e,e'\}$ of
non-adjacent edges, increase it by 1 once $e$ and $e'$ cross. 
Consider the set $S$ of 4 endpoints of $e$ and $e'$. In fact, $S$ corresponds 
to exactly 3 pairs of edges. If the convex hull of $S$ is a triangle, then 
none of these three pairs is crossing. If it is a quadrangle, then 1 of the 
three pairs is crossing and 2 are not. It follows that $\obf D$ does not exceed
the number of all possible $S$. This upper bound is attained if every $S$ has
a quadrangular hull, for instance, if the vertices of $D$ lie on a circle.
\item{$\obf{K_{s,t}}={s\choose2}{t\choose2}$.}
The upper bound is provable by the same argument as above, where a 4-point set
$S$ has 2 points in the $s$-point part of $V(D)$ and 2 points in the $t$-point
part. Such an $S$ corresponds to 2 pairs of non-adjacent edges, at most 1 of
which is crossing. This upper bound is attained if we put the two vertex parts
of $K_{s,t}$ on two parallel lines.
\item{$\obf{C_n}=n(n-3)/2$ if $n$ is odd.}
The value of $n(n-3)/2$ is attained by the $n$-pointed star drawing of $C_n$.
This is the maximum by a simple observation: $n(n-3)/2$ is the total number of
pairs of non-adjacent edges in $C_n$.
\end{description}

Let us state the upper bound argument we just used for the odd cycles in a general form.
Given a graph $G$ with $m$ edges, let
$$
\eps(G)={m\choose2}-\sum_{v\in V(G)}{\deg v\choose2}.
$$
Note that $\eps(G)=\frac12(m(m+1)-\sum_v\deg^2v)$, where the latter term is closely 
related to the variance of the vertex degrees. Since $\eps(G)$ is equal to the
number of pairs of non-adjacent edges in $G$, we have $\obf G\le\eps(G)$. Notice 
also a lower bound in terms of $\eps(G)$.

\begin{theorem}\label{thm:eps}
$\eps(G)/3\le\obf G\le\eps(G)$. Moreover, a drawing $D$ of $G$ with
$\obf D\ge\eps(G)/3$ is efficiently constructible.
\end{theorem}

\begin{proof}
Fix an arbitrary $n$-point set $V$ on a circle.
We use the probabilistic method to prove that there is a drawing $D$ with 
$V(D)=V$ having at least $\eps(G)/3$ edge crossings. Let $\mathbf D$ be a random
straight line embedding of $G$ with $V(\mathbf D)=V$, which is determined by a 
random map of $V(G)$ onto $V$. For each pair $e,e'$ of non-adjacent vertices of
$G$, we define a random variable $X_{e,e'}$ by $X_{e,e'}=1$ if $e$ and $e'$ cross
in $\mathbf D$ and $X_{e,e'}=0$ otherwise. Let $S$ be a 4-point subset of $V$. 
Under the condition that the set of endpoints of $e$ and $e'$ in $\mathbf D$ is
$S$, these edges cross one another in $\mathbf D$ with probability $1/3$. It 
follows that $X_{e,e'}=1$ with probability $1/3$. Note that 
$\obf{\mathbf D}=\sum_{\{e,e'\}}X_{e,e'}$. By linearity of the expectation, we have
$\EE{\obf{\mathbf D}}=\sum_{\{e,e'\}}\EE{X_{e,e'}}=\frac13\,\eps(G)$ and hence 
$\obf D\ge\frac13\,\eps(G)$ for at least one instance $D$ of $\mathbf D$. Such a $D$
is efficiently constructible by standard derandomization techniques, namely,
by the method of conditional expectations, see, e.g., \cite[Chapter 15]{ASp}.
\end{proof}

As a consequence of Theorem \ref{thm:eps}, we have $\obf G=\Theta(n^2)$ for a 
planar $G$ whenever $\delta(G)\ge2$ (the latter condition excludes the cases like
$\obf{K_{1,s}}=0$). Indeed, $\eps(G)<\frac92\,n^2$ because $m<3n$ for any planar 
graph. This bound is sharp in the sense that $\eps(G)\ge\frac92\,n^2-O(n)$ for 
maximal planar graphs of bounded vertex degree. A sharp lower bound for 
$\eps(G)$ is stated below.

\begin{theorem}\label{thm:epsn2}
$\eps(G)\ge\of{\frac{\delta(G)^2}8-o(1)} n^2$ for a planar graph $G$ with $\delta(G)\ge2$.
The constant $\delta(G)^2/8$ cannot be better here.
\end{theorem}

\begin{proof}
Let $A_k(G)=\setdef{v\in V(G)}{\deg v<k}$ and denote
$$a_k(G)=|A_k(G)|\quad\mbox{and}\quad
s_k(G)=\sum_{v\in V(G)\setminus A_k(G)}\deg v.
$$
West and Will \cite{WWi} prove that, if $k\ge12$,
then for every planar $G$ on $n\ge\frac32k-1$ vertices
we have
$$
a_k(G)\ge\frac{(k-8)n+16}{k-6}
$$
and
$$
s_k(G)<2\,n-16+\frac{12(n-8)}{k-6}.
$$
We begin with the bound
$$
\eps(G)>\frac12\of{m^2-\sum_{v\in V(G)}\deg^2v}.
$$
Set $\delta=\delta(G)$.
Let $\sigma=s_k(G)/n$ (to simplify the notation, we do not indicate the dependence
of $\sigma$ on $k$). Suppose that $k$ is large enough, namely, $k\ge14$.
Note that $0\le\sigma<2+12/(k-6)$. We now estimate $m$ from below and $\sum_{v}\deg^2v$
from above.
\begin{eqnarray*}
m=\frac12\sum_{v}\deg v=
\frac12\of{\sum_{v\in A_k(G)}\deg v+\sum_{v\notin A_k(G)}\deg v}&&\\
\ge\frac12\of{\delta(G)a_k(G)+s_k(G)}>\frac12\of{\frac{\delta(k-8)}{k-6}+\sigma}n.
\end{eqnarray*}
Furthermore,
$$
\sum_{v}\deg^2v=\sum_{v\in A_k(G)}\deg^2v+\sum_{v\notin A_k(G)}\deg^2v<
(k-1)^2n+f(\sigma)n^2,
$$
where
$$
f(\sigma)=\left\{
\begin{array}{rcl}
2+(\sigma-2)^2&\textrm{if}&2\le\sigma<2+12/(k-6),\\
1+(\sigma-1)^2&\textrm{if}&1\le\sigma<2,\\
\sigma^2&\textrm{if}&0\le\sigma<1.
\end{array}
\right.
$$
Thus,
$$
\eps(G)>g(\sigma)\,n^2-\frac{(k-1)^2}2\,n,\ \ \textrm{where}\ \ 
g(\sigma)=\frac12\of{\frac14\of{\frac{\delta(k-8)}{k-6}+\sigma}^2-f(\sigma)}.
$$
A routine calculation shows that
$$
\min\setdef{g(\sigma)}{0\le\sigma<2+\frac{12}{k-6}}=g(0)=
\frac{\delta^2}{8}\of{\frac{k-8}{k-6}}^2.
$$
We conclude that
$$
\eps(G)>\frac{\delta^2}{8}\of{\frac{k-8}{k-6}}^2n^2-\frac{(k-1)^2}2n>
\of{\frac{\delta^2}8-\frac{\delta^2}{2(k-6)}-\frac{(k-1)^2}{2n}}n^2
$$
whenever $k\ge14$ and $n\ge\frac32k-1$. Recall that $\delta(G)\le5$ for any planar $G$.
If we make $k$ a function of $n$ that grows to the infinity slower than $\sqrt n$,
then the factor in front of $n^2$ becomes $\delta^2/8-o(1)$ and 
we arrive at the claimed bound.

The optimality of the constant $\delta^2/8$ is ensured by regular planar graphs 
(i.e., cycles and cubic, quartic, and quintic planar graphs).
\end{proof}

As was already mentioned, for planar graphs we have $\obf G\le\eps(G)<\frac92\,n^2$,
where the bound for $\eps(G)$ cannot be improved. However, for $\obf G$ we can do
somewhat better.

\begin{theorem}\label{thm:obfn2}
$\obf G<3\,n^2$ for a planar graph $G$ on $n$ vertices.
\end{theorem}

\begin{proof}
Note that, if $K$ is a subgraph of $H$, then $\obf K\le\obf H$.
It therefore suffices to prove the theorem for the case that $G$ is
a maximal planar graph, that is, a triangulation.
Let $E$ be a (crossing-free, not necessary straight line) plane embedding
of $G$. Denote the number of triangular faces in $E$ by $t$ and note that
$3t=2m$. Based only
on facial triangles, let us estimate from below the number of non-crossing edge pairs in
an arbitrary straight line drawing $D$ of $G$. 
Let $P$ denote the set of all pairs of adjacent edges occurring
in facial triangles. Here we have $|P|=3t$ edge pairs which are non-crossing
in $D$. Furthermore, for each pair of edge-disjoint facial triangles $\{T,T'\}$ 
we take into account pairs of non-crossing 
edges $\{e,e'\}$ with $e$ from $T$ and $e'$ from $T'$. 
Since at most $3t/2$ pairs of facial triangles can
share an edge, there are at least ${t\choose2}-\frac{3t}2$ such $\{T,T'\}$.
We split this amount into two parts. Let $A$ consist of vertex-disjoint $\{T,T'\}$
and $B$ consist of $\{T,T'\}$ sharing one vertex.
As easily seen, every $\{T,T'\}$ in $A$ gives us at least 3 edge pairs $\{e,e'\}$ 
which are non-crossing in $D$. Every $\{T,T'\}$ in $B$ contributes
at least 2 pairs of non-adjacent edges and exactly 4 pairs of adjacent edges. 
However, 2 of the latter 4 edge pairs can participate in $P$.
We conclude that in $D$ there are at least $|P|+(3|A|+4|B|)/4$
non-crossing edge pairs. The factor of $1/4$ in the latter term is needed because 
an edge pair $\{e,e'\}$ can be contributed by 4 triangle pairs $\{T,T'\}$.
Thus, 
$$
\obf D\le{m\choose2}-3t-\frac34\of{{t\choose2}-\frac{3t}2}<
\frac12\,m^2-\frac38\,t^2=\frac13m^2.
$$
Since $m<3n$ as a simple consequence of
Euler's formula, we have $\obf D<3n^2$.
As $D$ is arbitrary, the bound for $\obf G$ follows.
\end{proof}

\section{Estimation of the shift complexity}\label{s:shift}

A basic fact about $\shift G$ is that this number is well defined.

\begin{theorem}[Wagner, F\'ary, Stein (see, e.g., \cite{NRa})]\label{thm:WFS}
Every planar graph $G$ has a straight line plane drawing.
In other words, $\shift G\le n-3$ if $n\ge3$.
\end{theorem}

If we seek for lower bounds, the following example is instructive
despite its simplicity: $\shift{mK_2}=m-1$.
It immediately follows that
$$
\shift G\ge\nu(G)-1.
$$

\begin{theorem}\label{thm:shift}
Let $G$ be a connected planar graph on $n$ vertices.
\begin{bfenumerate}
\item
If $\delta(G)\ge3$ (in particular, if $G$ is 3-connected) and $n\ge10$,
then $\shift G\ge(n-1)/3$.
\item
If $G$ is 4-connected, then $\shift G\ge(n-3)/2$.
\item
There is an infinite family of connected planar graphs $G$ with
$\delta(G)=2$ and $\shift G\le2$.
\end{bfenumerate}
\end{theorem}

\begin{proof}
Item 1 follows from the fact that, under the stated conditions on $G$,
we have $\nu(G)\ge(n+2)/3$ (Nishizeki-Baybars \cite{NBa}). Item 2 is true
because every 4-connected planar $G$ is Hamiltonian (Tutte \cite{Tut})
and hence $\nu(G)\ge(n-1)/2$ in this case. Item 3 is due to the bound
$\shift{K_{2,s}}\le2$. The latter follows from the elementary fact of plane geometry
stated in Lemma \ref{lem:geom} below.
\end{proof}

\begin{lemma}\label{lem:geom}
For any finite set of points $Z$ there are two points $x$ and $y$ such that
the segments with one endpoint in $\{x,y\}$ and the other in $Z$ do not
cross each other and have no inner points in $Z$.
\end{lemma}

\begin{proof}
Let $L$ denote the set of all lines going through at least two points
in $Z$. Fix the direction ``upward'' not in parallel to any line in $L$.
Pick up $x$ above every line in $L$ and $y$ below every line in $L$.
\end{proof}

The next question we address is this: How close is relationship between
$\shift G$ and $\nu(G)$? By Theorem \ref{thm:shift}, if $\delta(G)\ge3$ 
then both graph parameters are linear. However, if $\delta(G)\le2$,
the existence of a large matching is not the only cause of large
shift complexity.

\begin{theorem}\label{thm:shiftGs}
There is a planar graph $G_s$ on $3s+3$ vertices with $\delta(G_s)=2$
such that $\nu(G_s)=3$ and $\shift{G_s}\ge2s-6$.
\end{theorem}

\begin{figure}
\includegraphics{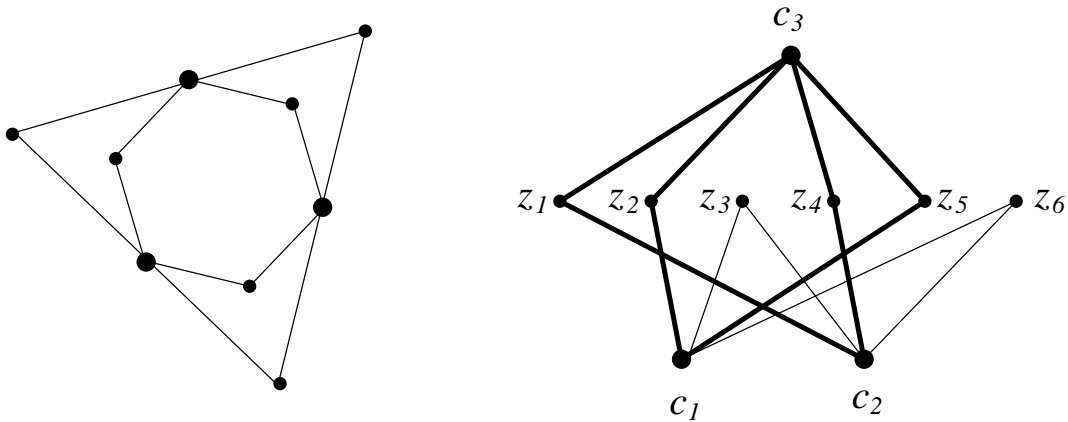}
\caption{$G_2$ and $F$ in $D_2$.}
\end{figure}

\begin{proof}
A suitable $G_s$ can be obtained as follows: take the multigraph
which is triangle with multiplicity of every edge $s$ and make it
graph by inserting a new vertex in each of the $3s$ edges (see Fig.~1).
Using Lemma \ref{lem:geom}, it is not hard to show that $\shift{G_s}\le2s+3$.
We now construct a drawing $D_s$ of $G_s$ with $\shift{D_s}\ge2s-6$.
Put vertices $z_1,\ldots,z_{3s}$ in this order in a line and the remaining vertices 
$c_0,c_1,c_2$ somewhere else in the plane.
Connect $z_i$ with $c_j$ iff $j\ne i\bmod3$. 
Therewith $D_s$ is specified. Denote the fragment
of $D_s$ induced on $\{z_1,z_2,z_4,z_5,c_0,c_1,c_2\}$ by $F$. It is not hard
to see that $F$ cannot be disentangled by moving only $c_0$, $c_1$, and $c_2$.
In fact, if in place of $z_1,z_2,z_4,z_5$ we take any quadruple $z_i,z_j,z_k,z_l$
with $i<j<k<l$, $i\equiv k\!\!\!\pmod 3$, and $j\equiv l\!\!\!\pmod 3$, this will give us
a fragment completely similar to $F$. To destroy all such fragments,
we need to move at least two vertices in every triple $z_{3h+1},z_{3h+2},z_{3h+3}$
($0\le h<s$) with possible exception for at most 3 of them. 
Therefore, making $2(s-3)$ shifts is unavoidable.
\end{proof}

Finally, we prove a complexity result.

\begin{theorem}\label{thm:hard}
Computing the shift complexity of a given drawing is an NP-hard problem. 
\end{theorem}

\begin{proof}
In fact, this hardness result is true even for drawings of graphs $mK_2$.
Given such a drawing $D$, consider its intersection graph $S_D$ whose
vertices are the edges of $D$ with $e$ and $e'$ adjacent in $S_D$ iff
they cross one another in $D$. Since computing the independence number
of intersection graphs of segments in the plane is known to be NP-hard
(Kratochv\'{\i}l-Ne\v{s}et\v{r}il \cite{KNe}), it suffices for us to
express $\alpha(S_D)$ as a simple function of $\shift D$.
Fix an optimal way of untangling $D$ and denote the set of edges whose
position was not changed by $E$. Clearly, $E$ is an independent set in $S_D$
and hence $\shift D\ge m-|E|\ge m-\alpha(S_D)$. On the other hand, $\shift D\le m-\alpha(S_D)$.
Indeed, fix an independent set $I$ in $S_D$ of the maximum size $\alpha(S_D)$.
Then $D$ can be untangled this way: we leave the edges in $I$ unchanged and shrink
each edge not in $I$ by shifting one endpoint sufficiently close to the other
endpoint. Thus, $\alpha(S_D)=m-\shift D$, as desired.
\end{proof}

\section{Concluding remarks and problems}\label{s:concl}
\mbox{}

\que
By Theorem \ref{thm:eps} we have $\frac13\,\eps(G)\le\obf G\le\eps(G)$.
The upper bound cannot be improved in general as $\obf{C_n}=\eps(C_n)$ for odd $n$.
Can one improve the factor of $\frac13$ in the lower bound?

\que
By Theorems \ref{thm:eps}, \ref{thm:epsn2}, and \ref{thm:obfn2} we have
$(\delta(G)^2/24-o(1))n^2\le\obf G\le 3\,n^2$ where $\delta(G)\ge2$ is necessary for
the lower bound. Optimize the factors in the left and the right hand sides.

\que
As follows from the proof of Theorem \ref{thm:eps}, there is an $n$-point set
$V$ (in fact, this can be an arbitrary set on the border of a convex body)
with the following property: Every graph $G$ of order $n$ has a drawing $D$
with $V(D)=V$ such that $\obf D\ge\frac13\obf G$. Can this uniformity result be
strengthened? Is there an $n$-point set $V$ on which one can attain $\obf D=\obf G$
for all $n$-vertex~$G$?

\que\label{oq:obfvsshift}
The following remarks show that the obfuscation
and the shift complexity of a drawing have, in general, rather
independent behavior.

\begin{description}
\item[\it Maximum $\obf D$ does not imply maximum $\shift D$.]
Consider $3K_{1,s}$, the un\-ion of 3 disjoint copies of the $s$-star.
It is not hard to imagine how a drawing attaining $\obf{3K_{1,s}}=3s^2$
should look (where every two non-adjacent edges cross) and 
it becomes clear that such a drawing can be untangled just by 2 shifts.
However, $\shift{3K_{1,s}}\ge s$ is provable similarly to Theorem \ref{thm:shiftGs}
(an upper bound $\shift{3K_{1,s}}\le s+2$ follows from Lemma \ref{lem:geom}).
\item[\it Maximum $\shift D$ does not imply maximum $\obf D$.]
The simplest example is given by a drawing of the disjoint union of $K_2$
and $K_{1,2}$ with only one edge crossing.
\item[\it Large $\obf D$ does not imply large $\shift D$.]
This can be shown by drawings of $\obf{K_{2,s}}$.
Indeed, we know that $\obf{K_{2,s}}={s\choose2}$ from Section \ref{s:obf} and
$\shift{K_{2,s}}\le2$ from Section \ref{s:shift} (the latter bound is exact if $s\ge4$).
\item[\it Large $\shift D$ does not imply large $\obf D$.]
Pach and Tardos \cite[Fig.~2]{PTa} show a drawing $D$ of the cycle $C_n$
with linear $\shift D$ and $\obf D=1$.
\end{description}

\que
In spite of the observation we just made that large $\obf D$ does not imply 
large $\shift D$, in some interesting cases it does. Pach and Solymosi \cite{PSo}
prove that every system $S$ of $m$ segments in the plane with $\Omega(m^2)$
crossings has two disjoint subsystems $S_1$ and $S_2$ with both $|S_1|=\Omega(m)$
and $|S_2|=\Omega(m)$ such that every segment in $S_1$ crosses all segments in $S_2$.
As $\shift S\ge\min\{|S_1|,|S_2|\}$, this result has an interesting consequence:
If $D$ is a drawing of $mK_2$ with $\obf D=\Omega(m^2)$, then $\shift D=\Omega(m)$.

\que
Theorem \ref{thm:hard} shows that computing $\shift D$ for a drawing $D$ of a graph $G$
can be hard even in the cases when computing $\shift G$ is easy. Is $\shift G$ hard
to compute in general? 
Theorem \ref{thm:eps} shows that $\obf G$ is polynomial-time approximable within
a factor of 3. Is exact computation of $\obf G$ NP-hard (Amin Coja-Oghlan)?

\subsection*{Acknowledgment}
I am thankful to the members of the `Algorithms and Complexity' group at the Humboldt
University of Berlin and to Taras Banakh for helpful discussions.
I thank  Sasha Ravsky for a simplification of the proof
of Theorem \ref{thm:obfn2} and an anonymous referee for useful comments,
in particular, for drawing my attention to the recent work done in \cite{Goaos,SWo,Bose}.

\end{document}